# Prolonged mixed phase induced by high-pressure in MnRuP

Raimundas Sereika[1,2,*], Wei Wu[3], Changyong Park[4], Curtis Kenney-Benson[4], Dale L. Brewe[5], Steve M. Heald[5], Jianbo Zhang[1], Sorb Yesudhas[1], Hongshan Deng[1], Bijuan Chen[1], Jianlin Luo[3], Yang Ding[1,*], Ho-kwang Mao[1,4,6].

[1]*Center for High Pressure Science and Technology Advanced Research, Beijing 100094, China*
[2]*Faculty of Science and Technology, Lithuanian University of Educational Sciences, Studentu 39, Vilnius 08106, Lithuania*
[3]*Beijing National Laboratory for Condensed Matter Physics and Institute of Physics, Chinese Academy of Sciences, Beijing 100190, China*
[4]*HPCAT, Geophysical Laboratory, Carnegie Institution of Washington, 9700 South Cass Avenue, Argonne, Illinois 60439, USA*
[5]*X-ray Science Division, Advanced Photon Source, Argonne National Laboratory, 9700 South Cass Avenue, Argonne, Illinois 60439, USA*
[6]*Geophysical Laboratory, Carnegie Institution of Washington, Washington DC 20015, USA.*

**Abstract.**
Hexagonally structured MnRuP was studied under high pressure up to 35 GPa from 5 to 300 K using synchrotron X-ray diffraction. We observed that a partial phase transition from hexagonal to orthorhombic symmetry started at 11 GPa. The new and denser orthorhombic phase coexisted with its parent phase for an unusually long pressure range, $\Delta P \approx 50$ GPa. We attribute this structural transformation to a magnetic origin, where a decisive criterion for the boundary of the mixed phase lays in the different distances between the Mn-Mn atoms. In addition, our theoretical study shows that the orthorhombic phase of MnRuP remains steady even at very high pressures up to ~ 250 GPa, when it should transform to a new tetragonal phase.

**Keywords:** MnRuP; high-pressure; structure prediction; X-ray diffraction; mixed phases.

*Corresponding authors:
raimundas.sereika@hpstar.ac.cn; yang.ding@hpstar.ac.cn

# I. INTRODUCTION

Ternary phosphide MnRuP belongs to the well-known family of MM'X compounds (M = Mn, Cr; M' = Ru, Rh, Pd metal and X = As, P). It is an incommensurate antiferromagnetic metal that crystallizes in the non-centrosymmetric $Fe_2P$-type crystal structure. MnRuP magnetic ac susceptibility, heat capacity, and neutron diffraction data confirmed three magnetic transitions at low temperatures [1, 2]. These discoveries were made a long time ago, and since then, only a few studies have been conducted on magnetization issues at ambient pressure [3, 4]. The recent discovery of a MnP superconductor, a rare case of a noncollinear helimagnetic superconductor under high pressure, generated great interest in understanding microscopic magnetic properties and their interplay with superconductivity in MnP-type materials and similar systems [5-7]. MnRuP has many advantages for showing interesting properties under high pressure because lots of ternary transition metal phosphides with an ordered $Fe_2P$-type hexagonal structure are high-temperature superconductors and their crystallographic ordering is reported to be highly sensitive to external parameters [8, 9]. Very recent research shows that below the *Neél* transition at 250 K, MnRuP exhibits hysteretic anomalies in its resistivity and magnetic susceptibility curves as the propagation vectors of the spiral spin structure change discontinuously across $T_1 = 180$ K and $T_2 = 100$ K [10]. Temperature-dependent X-ray diffraction data indicates that the first-order spin reorientation occurs in the absence of a structural transition. However, no study on MnRuP under high pressure has been reported so far.

Here, we identify a new pressure-induced phase of MnRuP that evolves slowly during compression. The other MM'X family compounds (e.g., MnRhP, MnRhAs) also manifest structural transitions, which have not been addressed to date but clearly show detuned mixed phase behavior because the X-ray diffraction peaks from the original phase coexist with incoming new phase peaks for quite a broad span of pressures [11]. For instance MnRhP have mixed phase from 34 to 48 GPa ($\Delta P = 14$ GPa), and MnRhAs – from 26 to 59.6 GPa ($\Delta P = 33.6$

GPa). In this regard, MnRuP has the lowest starting pressure – 11 GPa and potentially the longest two-phase persistence known to date for intermetallic compounds. Therefore, in this work we focus on the structural transition of MnRuP with its unusual symmetry exchange mechanism at various pressure-temperature conditions.

## II. METHODS

The MnRuP crystals were grown using a Sn-flux method. The starting materials were Mn (Cerac, powder, 99.9%), Ru (Cerac, powder, 99.9%), P (Alfa Aesar, powder, 99.99%), and Sn (Cerac, shot, 99.99%). All of the manipulations were completed in an Argon-filled glove box with moisture and oxygen levels of less than 1 ppm. The materials with an atomic ratio of Mn:Ru:P:Sn = 1:1:1.05:30 were added to an alumina crucible, which was placed in a quartz ampoule, and subsequently sealed under a reduced pressure of $10^{-4}$ Torr. The quartz ampoule was heated up to 650°C for 10 h and maintained for a period of 8 h, then heated up to 1000°C for 15 h, maintained for 6 h, and slowly cooled down to 600°C for 50 h. At this temperature, the liquid Sn flux was filtered. The prepared samples were washed further in an ultrasonic bath several times to make sure no contamination remained in the samples.

At ambient pressure, MnRuP adopts a hexagonal lattice (space group $P\bar{6}2m$) and unit cell parameters of $a = b = 6.257$ Å and $c = 3.523$ Å [1]. To date, the atomic positions have not been reported in the literature, but the implicit locations for the Mn atoms are at 3$g$, Ru at 3$f$, and P within the 2$c$ and 1$b$ positions [8]. Our synchrotron angle-dispersive X-ray diffraction (ADXRD) results of the finely-prepared powders from single MnRuP crystals are in good agreement with previously reported lattice parameters and the predicted positions of the atoms. The high-pressure experiments were performed using a Mao-type symmetric diamond anvil cell where the neon gas and silicon oil served as the pressure-transmitting medium for the X-ray diffraction and X-ray absorption measurements, respectively. The solved structural information and detailed sample preparation procedure for high-pressure measurements can be found in the Supplemental Material [12].

### III. RESULTS

#### A. Experimental evidence

Under high pressure, it is expected that the system will transform to an orthorhombic TiFeSi-type ($Co_2P$-type) structure (space group *Pnma*) or a tetragonal $Fe_2As$-type one (*P4/nmm*) because both phases have cohesive energies close to the $Fe_2P$-type structure [11]. This idea is also supported by the fact that some MM'X compounds crystallize in the mentioned phases [3, 4, 8] and such phase transformations have been detected in similar systems when the temperature was varied, or doping methods used [13-16]. Our high-pressure diffraction data clearly showed that the new phase appears at 11 GPa (see Fig. 1). The indexing of the ADXRD patterns for new peaks gave the highest figure of merit for the monoclinic ($P2_1$) phase rather than others. However, our theoretical calculations on structure prediction revealed that such a phase is energetically away from the original phase and its existence in this pressure range is unlikely. The new phase met expectations for an orthorhombic TiFeSi-type structure with the *Pnma* space group and agrees well with our theoretical prediction, fitting well with only negligible errors. Thus, we conclude that the mixed phase starting at 11 GPa consists of hexagonal ($P\bar{6}2m$) and orthorhombic (*Pnma*) contributions. Figures 1 (a) and 1 (b) show the comparison of the high-pressure ADXRD data before the transition at 8.5 GPa (hexagonal phase alone) and post-transition at 11.5 GPa (a mixture of the hexagonal and orthorhombic phases). The Le Bail refinement for the mixed phase at 11.5 GPa gave hexagonal unit cell parameters of $a = b = 6.160$ Å, $c = 3.399$ Å, and orthorhombic unit cell parameters of $a = 6.019$ Å, $b = 4.186$ Å, and $c = 7.143$ Å. The complete list of lattice parameters for both phases under high-pressure is given in the Supplemental Material [12].

For the hexagonal phase, the axial ratio, $c_h/a_h$, and unit-cell volume versus pressure changes the slope at the transition point. The pressure coefficient of the axial ratio, $d(c_h/a_h)/dP$, was determined to be $-1.012 \times 10^{-3}$ GPa$^{-1}$. This value is within the range reported for MM'X ternary systems in Ref [11]. The pressure dependence of the unit-cell volume per molecule and $d$-spacings of the diffraction peaks plotted as a function of pressure are presented in Figs. 1(d) and 1(e), respectively. The relationships between the volume and pressure for both phases were fitted to the Birch-Murnaghan third-order equation of state using EoSFit software [17]. The best-fit yielded the bulk modulus, $K_0 = 158.0$ GPa, with its fixed derivative value $K_0' = 4$ for the low-pressure phase. The value of the bulk modulus is in between what was found for MnRhAs ($K_0 = 117.2$ GPa, $K_0' = 4$) and MnRhP ($K_0 = 213.5$ GPa, $K_0' = 4$) [11]. The high-pressure phase gives an increased value of $K_0 = 178.4$ GPa. A good fitting was only possible with a fixed derivative value of $K_0' = 2$. The new orthorhombic phase brings a 3.35 % reduction in volume and ~ 3.7 % increase in density, compared with the hexagonal phase. The volume collapse and density change is not as big compared to what was found in some other manganese compounds under high-pressure: e.g., the manganese chalcogenides (MnS, MnSe) [18] and mineral hauerite (MnS$_2$) [19]. However, considering the similarity to MnS$_2$ the high-pressure transition to a mixed phase at 11 GPa could also be driven by the magnetic mechanism [19].

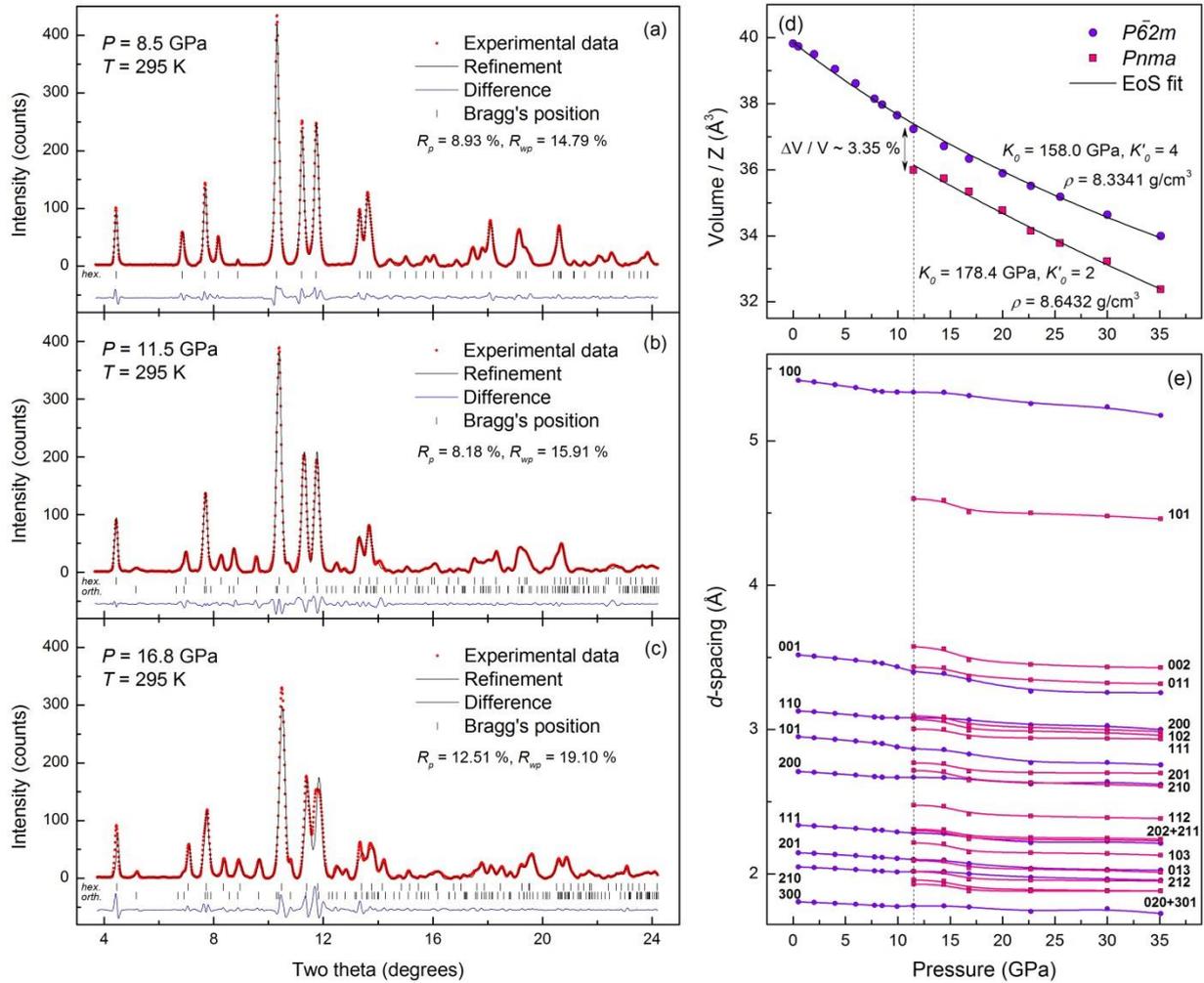

**Figure 1.** The fitting of the ADXRD data for MnRuP powders at room temperature and different pressures. The wavelength of the incident X-ray beam is 0.4133 Å. (a) The data collected before the phase transition at 8.5 GPa. (b) and (c) The data collected after the transition at 11.5 GPa and 16.8 GPa, respectively. (d) Volumes per formula unit as a function of pressure indicating a new MnRuP phase appearance at 11 GPa (dashed line). The symbols are the experimental data: purple points indicate the hexagonal phase, and pink squares denote the orthorhombic phase. The solid lines are the calculated third-order Birch−Murnaghan equation of state (EoS) fit to the experimental data. (e) Pressure dependence of *d*-spacing.

The MnRuP was probed for a possible Mn valence state during the structural transition using an X-ray absorption near edge structure (XANES) technique at the Mn *K*-edge (~6.54 keV). XANES provides the element specific formal valence and information on the chemical and electronic structures including the coordination environment. The Mn *K*-edge XANES spectra of the reference compounds and the MnRuP sample at ambient conditions are shown in Fig. 2a. The Mn compounds generally have a single pre-edge peak around 6542 eV in the XANES spectra, indicating that the Mn atoms occupy sites without a center of inversion. The electronic characteristics of the Mn atoms in each sample can be obtained by analyzing the Mn *K*-edge shift in the XANES spectra. In principle, it is possible to obtain a quantitative estimation, averaged for all the Mn atoms in the sample, for the Mn oxidation state. Each different chemical species of Mn contributes its specific weight to the experimental spectrum. The energy shifts on the absorption edge are directly related to the average oxidation state of the absorbent atom [20-22]. The absorption edge corresponding to $Mn^{3+}$ is at smaller energies than the corresponding one for $Mn^{4+}$. The Mn *K*-edge in MnRuP is very similar to the spectrum of Mn-metal and is comparable in shape to the Mn *K*-edge in MnP rather than the manganese oxides (Fig. 2a). MnP is known to have a $Mn^{3+}$ valence state and it is assumed that the three 3*d*-electrons of $Mn^{3+}$ are spin-up and one electron is spin-down [23]. Based on the very close similarity between the Mn *K*-edge spectra in the MnRuP and MnP compounds, it is fair to assume that the manganese in MnRuP also has a $3^+$ valence state. In fact, this is true because the Mn absorption edge energy value of ~ 6548 eV in MnRuP is attributed to $Mn^{3+}$ [24-26]. Therefore, it is clear that MnRuP has a different charge distribution to the same class ZrRuP, whose Zr oxidation state is $4^+$ [27], suggesting that the bonding of ZrRuP can be described in terms of the oxidation states $Zr^{4+}(RuP)^{4-}$. Therefore, the bonding in for MnRuP could be written as $Mn^{3+}(RuP)^{3-}$ using the same concept.

The XANES spectra measured at different pressures were normalized to a unit edge jump to account for possible variations in the sample thickness as the pressure increased. Their energy derivatives are shown in Fig. 2b. The spectra do not show any shift in the Mn absorption edge as pressure is applied. This indicates that the structural transition was not accompanied by a change in the Mn valence state and therefore, a change in the Ru valence state is also unlikely.

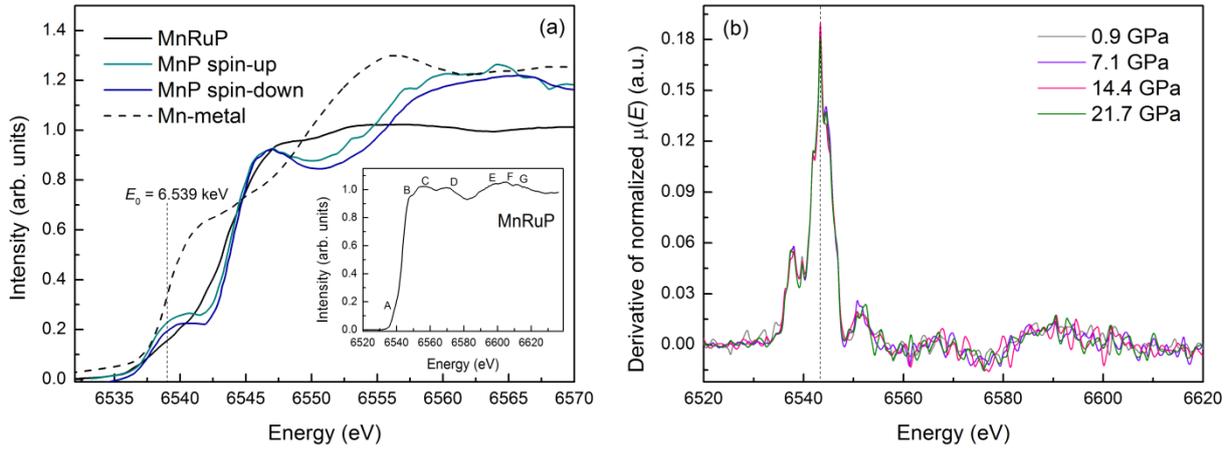

**Figure 2.** (a) The comparison of the Mn *K*-edge XANES spectra for selected manganese materials at ambient pressure and room temperature. The spectra of the Mn-metal and MnP taken from [21, 28] compared to the MnRuP data. The inset shows the Mn *K*-edge spectrum for MnRuP alone, with its features marked by a sequence of letters. (b) An expanded view of the energy derivative of the MnRuP XANES spectra, showing the absence of any detectable Mn valence transition in the given pressure range.

To describe the phase exchange process in detail, we evaluated the phase weight fraction data because the coexistence of the diffraction peaks from the original hexagonal phase was evident over the wide range of pressures studied (11 – 35 GPa). Our fitting results show a gradual decrease of the hexagonal and an increase of the orthorhombic phase contributions in the mentioned range of pressures (see Fig. 3). This suggests that the phase transition may be local rather than global. Assuming the variation has a linear course, the hexagonal and orthorhombic contributions should intersect at 36 GPa. At this pressure point, the two phases have an equal weight fraction value. Extrapolated fitting results suggest that the transition could be extended to a maximum of up to ~ 61 GPa. In this case, the mixed phase should cover the $\Delta P = 50$ GPa range of pressure. In addition, our low-temperature studies revealed that the transition site remains unaffected by the temperature change from 5 to 300 K within a small ~ 1 GPa error, which may occur due to pressure measurement inaccuracies at low temperatures.

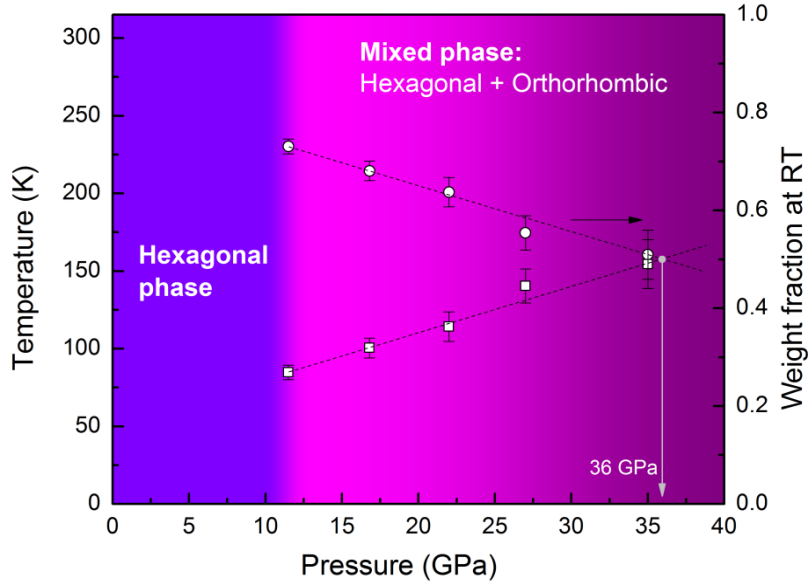

**Figure 3.** The pressure dependence of the temperature and weight fraction in MnRuP. The symbols represent the data extracted by two-phase fitting. The round and square symbols belong to the hexagonal and orthorhombic phases, respectively. The dashed curves are linear fits for the hexagonal and orthorhombic phase data.

### B. Theoretical predictions

All possible stable and metastable phases in the MnRuP system were searched for using the evolutionary algorithm as implemented in USPEX software [29-31]. A series of structures were obtained and the lowest enthalpy structures were considered. The calculation indicated three phases within a 0.07 eV/f.u. range from the original hexagonal ($P\bar{6}2m$): orthorhombic (*Pnma*), tetragonal (*P4/nmm*) and monoclinic (*Pm*). These phases are energetically very close to each other at ambient pressure and thus, possibly synthesizable. However, for *P4/nmm* and *Pm* the increase of pressure dictates a strong deviation from the lowest enthalpy $P\bar{6}2m$ and *Pnma* phases. Conversely, the *Pnma* is very stable under high pressure and do not change much. According to our theoretical prediction, the next stable phase will appear only at very high pressures ≥ 250 GPa. At these pressures, the structure for MnRuP is predicted to be tetragonal (*P4mm*). (see calculation details and generated structural information provided in [12]). Figure 4 summarizes the experimental and theoretical investigation on the high-pressure phases of MnRuP.

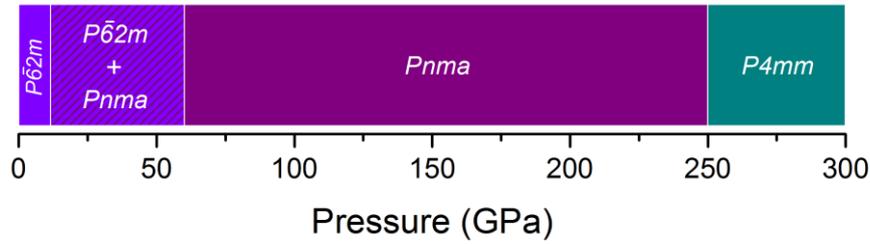

**Figure 4.** Crystal symmetry phase diagram of MnRuP in the pressure range from 0 to 300 GPa.

## IV. DISCUSSION

Refs. [11, 32] discuss and show that a decrease of the interatomic distances between the first-nearest neighboring (1NN) Mn atoms lying on the same *c*-plane of the hexagonal cell can cause a change of the magnetic order in such systems. It is known that at a particular Mn–Mn distance (~ 3.0 Å) the effective exchange interaction coefficient between the Mn atoms turns from positive to negative [33]. In this study, we do not have experimental evidence of the magnetic origin, but in view of past research on the MM'X family compounds, the transition in MnRuP should definitely have a magnetic origin with the same mechanism observed in MnRhAs [9, 34-37]. This assumption is based on the very close structural similarities between MnRuP and MnRhAs, as well as several important facts. Firstly, the original compound, $Fe_2P$ itself, is well known to have a magnetic phase transition under high pressure [38]. Secondly, the magnetic properties in these materials have proven to be very sensitive to external parameters. The magnetic structure is considered to be strongly dependent on the lattice constants because shrinking of these lattice parameters causes a phase transformation from an antiferromagnetic to a ferromagnetic state using both external pressure [34] or chemical pressure [9, 35]. Therefore, the structural changes in MnRuP at 11 GPa can be considered the beginning of a pressure-induced antiferromagnetic-to-ferromagnetic transition lead by the Ruderman–Kittel–Kasuya–Yosida (RKKY) interactions, where the effective exchange parameters of pair-wise metal–metal magnetic couplings plays a crucial role [36, 37]. This mechanism agrees well with the formation of the mixed phase. Moreover, our *ab initio* investigation confirms that the hexagonal phase is antiferromagnetic with high-spin among the Mn atoms, where each plane of the 1NN Mn atoms has a differently oriented spin. The orthorhombic phase was ferromagnetic, as we predicted.

The most appropriate explanation for why this compound enters into a mixed phase instead of a single phase is that the transition depends not only on particular 1NN Mn distances but also on the distances between the second nearest neighboring (2NN) Mn atoms lying on the neighboring *c*-planes. When the distance of the 1NN Mn atoms reaches a critical value (assumed to be 3.0 Å), part of the sample enters into a new phase, while the rest remains in the old phase, because acting forces are not strong enough to convert all the matter into the new phase right away and the distance of 2NN is well behind 1NN (see Fig. 5). When the distance of the 2NN Mn atoms in the remaining hexagonal lattices reaches the critical value then all the matter will transform into the new phase. Thus, one can define the completion of the mixed phase transition by the lattice parameter $c_h$, because the distance 2NN = $c_h$. This value for MnRuP at 35 GPa is $c_h$ = 3.2 Å. The extrapolation of the lattice parameter course $c_h$ *versus* pressure predicts that the $c_h$ = 3.0 Å value can be reached at around 60 GPa. This agrees well with our weight fraction analysis data and coincides with the $P\bar{6}2m$ – $Pnma$ mixed phenomena under high pressure.

It is very likely that other hexagonal MM'X family compounds have a similar phase exchange process when pressure is applied. However, MnRuP is more favorable to explore the mixed phase behavior in this process sequence since its phase transition starts at a much lower pressure of − 11 GPa than other Mn-M'X compounds. In comparison, the diffraction peaks of the new high-pressure phase for MnRhAs only begins to appear at 26 GPa. Therefore, it is unsurprising that the high-pressure phase in MnRhAs remained unsolved in Ref. [11]. The continuous high-pressure mixed phase often limits structural analysis.

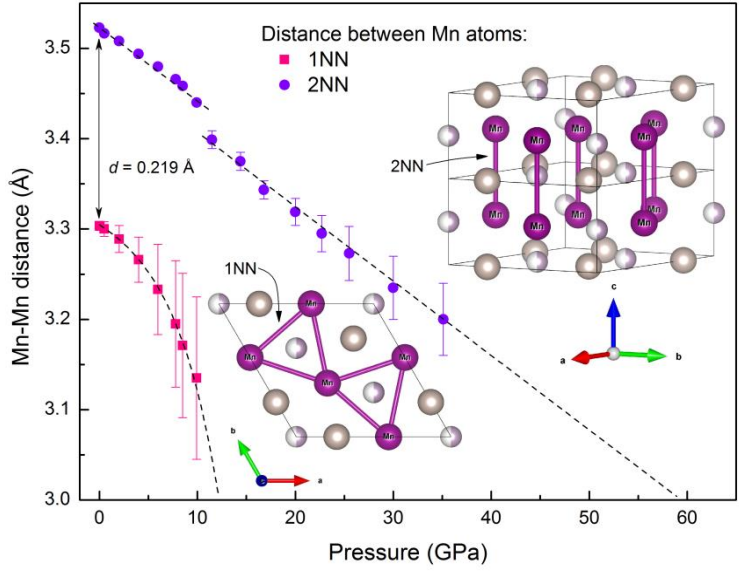

**Figure 5.** Pressure dependence of the first-nearest neighboring (1NN) and second nearest neighboring (2NN) Mn–Mn distances in the low-pressure phase of MnRuP. The 1NN distance decreases non-linearly while the 2NN distance shows a linear decrease with increasing pressure. The dashed lines are guides to the eye.

## V. CONCLUSION

In conclusion, our experimental ADXRD high-pressure and low-temperature studies revealed that a new structural phase transition of MnRuP took place at 11 GPa and remained unaffected by the temperature change from 5 to 300 K. The transition from a pure hexagonal to a mixed (hexagonal, $P\bar{6}2m$ and orthorhombic, $Pnma$) phase maintained the initial atomic oxidation states. The possible transition mechanism likely had a magnetic origin, initiated by the shortened distances between the 1NN Mn-Mn atoms and lead by the RKKY interactions. However, the inception was not strong enough to convert the entire sample into the orthorhombic phase and, thus, the compression above 11 GPa resulted in a mutually slow decrease of the hexagonal and increase of the orthorhombic phase contributions. The mixed phase maintained the hexagonal phase in the experimentally studied pressures up to 35 GPa. The boundary of the prolonged mixed phase was assigned to the 2NN Mn-Mn distances, which can presumably terminate mixed behavior at ~ 60 GPa, resulting in one of longest mixed phase ranges for intermetallic compounds of $\Delta P \approx 50$ GPa. The calculations using conventional structure prediction methods supports the *Pnma* as the most stable phase up to 250 GPa and indicates that MnRuP transforms to a new tetragonal phase *P4mm* above 250 GPa.

**Acknowledgements**

The angle-dispersive x-ray diffraction measurements were performed at HPCAT (Sector 16-BM-D) and x-ray-absorption experiments were performed at XSD (Sector 20-BM-B) of the Advanced Photon Source, a US Department of Energy (DOE) Office of Science user facility operated by Argonne National Laboratory (ANL) under Contract No. DE-AC02-06CH11357. HPCAT operation is supported by DOE-NNSA under Award No. DE-NA0001974, with partial instrumentation funding by NSF. XSD operation is supported by the DOE and the Canadian Light Source. C.P. and C.K.B acknowledge the support of DOE-BES/DMSE under Award DE-FG02-99ER45775. Y.D and H.-k.M. acknowledges the support from DOE-BES under Award No. DE-FG02-99ER45775 and NSFC Grant No. U1530402. This work is also supported by National Key R&D Program of China 2018YFA0305703 and Science Challenge Project, No TZ2016001.

# Supplemental Material

**Experimental details.**

High-pressure X-ray diffraction (XRD) experiments were performed using a Mao-type symmetric diamond anvil cell (DAC) with 400 μm culet anvils. A stainless steel gasket was precompressed to a 35 μm thickness, and a hole of 150 μm was drilled to load the sample, a ruby for pressure determination [1], and neon gas to serve as the pressure-transmitting medium [2]. The *in situ* high-pressure XRD measurements were carried out in the angle-dispersive mode at beamline 16-BM-D of the Advanced Photon Source (APS), Argonne National Laboratory. The incident monochromatic X-ray beam energy was set to 30.0 keV ($\lambda$ = 0.4133 Å), where the sample-detector distance was 310.91 mm. The wavelength of the X-ray was periodically calibrated using a $CeO_2$ standard. Diffraction patterns were recorded on a MAR345 image plate and then integrated using DIOPTAS software [3]. Indexing was carried out in EXPO2014 [4]. The refinements were performed using Jana2006 [5].

High-pressure XAS experiments were performed on manganese by investigating the X-ray absorption near edge structure (XANES) at beamline 20-BM-B of APS. A panoramic DAC with 400 μm diamonds was used to collect the spectra at the *K*-edge. XANES measurements were performed in fluorescence geometry, where the X-ray beam travels through a beryllium gasket, to avoid contamination of the XANES spectra by the Bragg peaks from the diamond anvils. The gasket was initially precompressed to ~70 μm, and then the whole culet area was drilled and replaced by boron nitride powder, which was compressed again to form a 60 μm radius hole drilled at the center of the boron nitride insert. Then, the sample, a ruby sphere, and silicon oil as a pressure medium were all loaded into the prepared hole. The collected data (Fig. 1) was processed and analyzed using programs from the Demeter package [6].

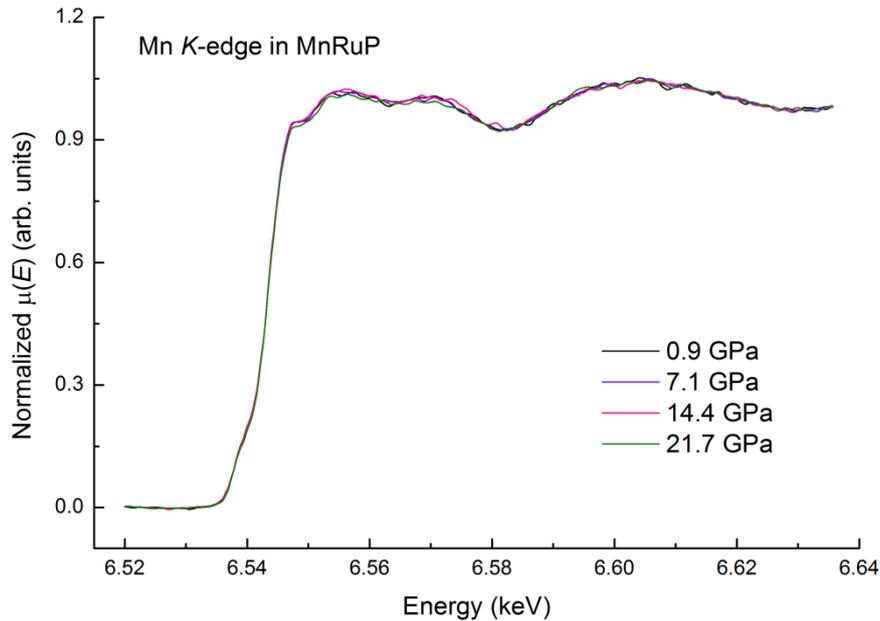

**Figure 1.** Normalized Mn *K*-edge XANES spectra of MnRuP at room temperature.

**Calculation details and generated structural information.**

The *ab initio* structural relaxations were carried out using density functional theory (DFT) with functionals from the VASP package [7, 8]. We selected the Perdew-Burke-Ernzerhof generalized gradient (PBE-GGA) and local density (LDA) approximations. The plane–wave kinetic energy cutoff was set to 320 eV and a Brillouin zone sampling resolution $2\pi \times 0.06$ Å$^{-1}$ was used. For comparison and verification, PBE-GGA with spin-orbit interaction (PBE+so) was also used in this study. In this case, a more precise resolution was guaranteed by setting the cutoff to 500 eV and the Brillouin zone sampling to $2\pi \times 0.04$ Å$^{-1}$. All structures were relaxed at constant pressure and 0 K, and the enthalpy was used as fitness.

Figure 2 shows the relative enthalpy diagram of the most energetically favorable phases up to 275 GPa. Here, the *P$\bar{6}$2m* and *Pnma* run almost parallel for all the pressures calculated. Surprisingly, the *Pnma* is a very stable phase of MnRuP, which maintains stability for a long pressure range. Our investigation on *Pnma* shows that different functionals render a similarly linear course to the *P6̄2m*. A small value of ΔH is a clear indication that the boundaries of these phases are not far from each other. The *ab initio* calculation using PBE with the spin-orbit interaction slightly adjusted the values of ΔH and the use of more appropriate functionals with different parameters may give better values as well. However, it is clear that the relative enthalpy of the *Pnma* phase never crossed or touched the *P6̄2m*, which could cause an immediate transition to a single phase.

The data provided in Table 1a is the structural and atomic information for the lowest energy structures (*P6̄2m*, *Pnma*, *P4/nmm*, and *Pm*) as obtained from the USPEX software at 11 GPa by using the PBE-GGA functional with spin-orbit interaction. The data in Table 1b is generated for the most stable structure at 250 GPa. The accompanying side pictures (produced by VESTA [9]) show the atomic arrangement in the unit cell.

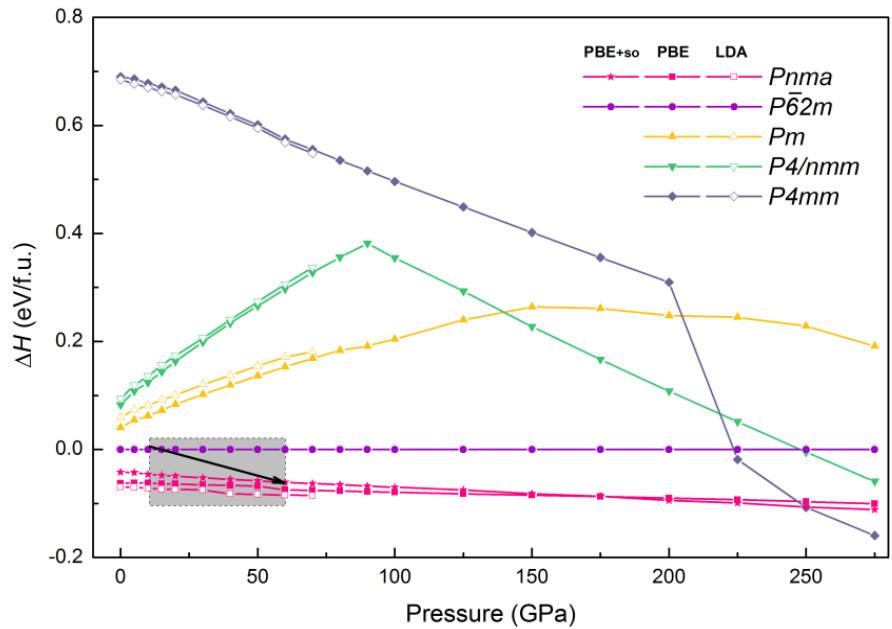

**Figure 2.** The computed relative enthalpy diagram as a function of pressure for MnRuP. The enthalpy of the *P6̄2m* phase was set to zero for every pressure as a reference. Here, the solid and empty points represent the results obtained using the PBE and LDA functionals, respectively. The rectangle with the dotted edges separates an area where the mixed phase was experimentally observed. The arrow indicates the transition course from *P6̄2m* to *Pnma*. Note that the higher energy symmetries were excluded from the diagram because they do not influence the hexagonal MnRuP or are unstable.

**Table 1a.** Crystallographic information on the structures predicted at 11 GPa using the PBE+so functional.

| | *Monoclinic Pm (no. 6),* $\alpha = 90$, $\beta = 90.3211$, $\gamma = 90$. $a = 5.7892$, $b = 3.7397$, $c = 6.6595$. | | |
|---|---|---|---|
| **Atom** | *x* | *y* | *z* |
| **Mn1** | 0.87998 | 0.50000 | 0.68104 |
| **Mn2** | 0.59019 | 0.00000 | 0.81663 |
| **Mn3** | 0.10220 | 0.00000 | 0.17707 |
| **Mn4** | 0.76501 | 0.50000 | 0.07889 |
| **P1** | 0.96364 | 0.00000 | 0.87099 |
| **P2** | 0.50200 | 0.50000 | 0.64697 |
| **P3** | 0.99076 | 0.50000 | 0.35722 |
| **P4** | 0.48999 | 0.00000 | 0.12908 |
| **Ru1** | 0.20930 | 0.00000 | 0.59317 |
| **Ru2** | 0.24741 | 0.50000 | 0.91343 |
| **Ru3** | 0.38579 | 0.50000 | 0.32187 |
| **Ru4** | 0.73645 | 0.00000 | 0.41353 |

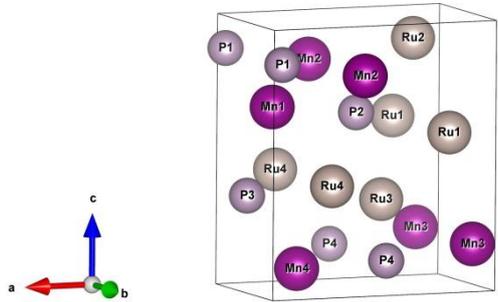

| | *Tetragonal P4/nmm (no. 129),* $\alpha = 90$, $\beta = 90$, $\gamma = 90$. $a = 3.5838$, $b = 3.5838$, $c = 5.6628$ | | |
|---|---|---|---|
| **Atom** | *x* | *y* | *z* |
| **Mn** | 0.50000 | 0.50000 | 0.00000 |
| **Ru** | 0.50000 | 0.00000 | 0.64980 |
| **P** | 0.50000 | 0.00000 | 0.24684 |

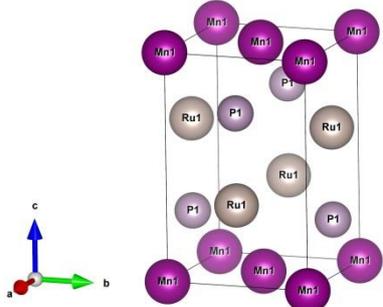

| | *Orthorhombic Pnma (no. 62),* $\alpha = 90$, $\beta = 90$, $\gamma = 90$. $a = 5.7781$, $b = 3.7306$, $c = 6.6167$. | | |
|---|---|---|---|
| **Atom** | *x* | *y* | *z* |
| **Mn** | -0.64309 | 0.25000 | -0.43077 |
| **Ru** | 0.01607 | 0.75000 | -0.33614 |
| **P** | -0.73452 | 0.75000 | -0.61537 |

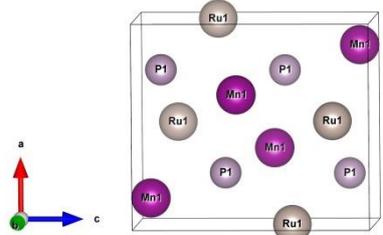

| | *Hexagonal P-62m (no. 189),* $\alpha = 90$, $\beta = 90$, $\gamma = 120$. $a = b = 5.7484$, $c = 3.7493$. | | |
|---|---|---|---|
| **Atom** | *x* | *y* | *z* |
| **Mn** | 0.74650 | 1.00000 | 0.50000 |
| **Ru** | 0.40357 | 1.00000 | 1.00000 |
| **P1** | 1.00000 | 1.00000 | 1.00000 |
| **P2** | 0.66667 | 0.33333 | 0.50000 |

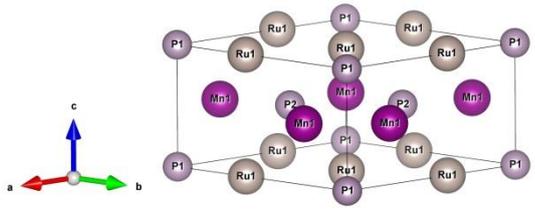

**Table 1b.** Crystallographic information on the most stable structure at 250 GPa.

| | Tetragonal P4mm (no. 99), $\alpha$ = 90, $\beta$ = 90, $\gamma$ = 90. $a$ = 2.414, $b$ = 2.414, $c$ = 8.0211. | | |
|---|---|---|---|
| **Atom** | *x* | *y* | *z* |
| **Mn1** | 0.00000 | 0.00000 | 0.71053 |
| **Mn2** | 0.00000 | 0.00000 | 0.41708 |
| **P1** | 0.50000 | 0.50000 | 0.27430 |
| **P2** | 0.50000 | 0.50000 | 0.57034 |
| **Ru1** | 0.00000 | 0.00000 | 0.10693 |
| **Ru2** | 0.50000 | 0.50000 | 0.89658 |

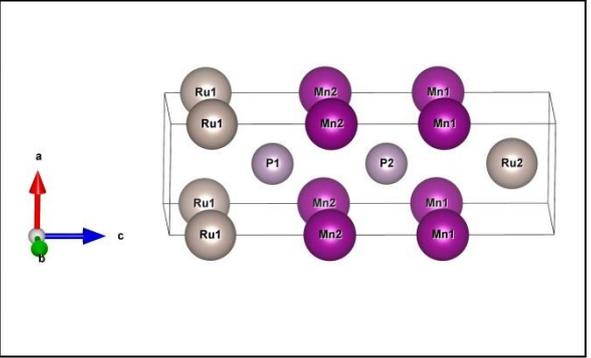

## Refinement details of the hexagonal and orthorhombic phases under compression.

**Table 2.** The change of the MnRuP hexagonal (*P-62m*) unit cell.

| *P*, GPa | *a*, Å | *c*, Å | *c/a* | *V*, Å$^3$ |
|---|---|---|---|---|
| **0.0001** | 6.257 | 3.523 | 0.563 | 119.45 |
| **0.50** | 6.256 | 3.517 | 0.562 | 119.21 |
| **2.01** | 6.245 | 3.508 | 0.562 | 118.49 |
| **4.00** | 6.222 | 3.494 | 0.562 | 117.14 |
| **5.98** | 6.200 | 3.480 | 0.561 | 115.85 |
| **7.78** | 6.174 | 3.466 | 0.561 | 114.43 |
| **8.50** | 6.167 | 3.459 | 0.561 | 113.91 |
| **9.93** | 6.157 | 3.440 | 0.559 | 112.93 |
| **11.50** | 6.160 | 3.399 | 0.552 | 111.70 |
| **14.40** | 6.139 | 3.375 | 0.550 | 110.14 |
| **16.80** | 6.135 | 3.343 | 0.545 | 108.99 |
| **22.70** | 6.110 | 3.295 | 0.539 | 106.53 |
| **30.00** | 6.090 | 3.235 | 0.531 | 103.91 |
| **35.10** | 6.066 | 3.200 | 0.528 | 101.97 |

**Table 3.** The change of the MnRuP orthorhombic (*Pnma*) unit cell.

| *P*, GPa | *a*, Å | *b*, Å | *c*, Å | *V*, Å$^3$ |
|---|---|---|---|---|
| **11.50** | 6.019 | 4.186 | 7.143 | 179.96 |
| **14.40** | 6.010 | 4.177 | 7.118 | 178.68 |
| **16.80** | 5.999 | 4.163 | 7.076 | 176.69 |
| **22.70** | 5.944 | 4.104 | 6.998 | 170.73 |
| **30.00** | 5.900 | 4.064 | 6.930 | 166.15 |
| **35.10** | 5.870 | 4.021 | 6.858 | 161.88 |

**Table 4.** The atomic positions of the hexagonal unit cell at ambient pressure and room temperature as refined using the Rietveld method.

| Atom | *x* | *y* | *z* | occ. |
|---|---|---|---|---|
| **Mn** | 0.59667 | 0.00000 | 0.50000 | 1.0 |
| **Ru** | 0.26103 | 0.00000 | 0.00000 | 1.0 |
| **P1** | 0.33333 | 0.66667 | 0.00000 | 0.5 |
| **P2** | 0.00000 | 0.00000 | 0.50000 | 0.5 |